\documentclass[aps,prl,twocolumn,superscriptaddress,nofootinbib]{revtex4-1}
\usepackage{amsmath}
\usepackage{graphicx}
\usepackage{bm}
\usepackage[normalem]{ulem}
\usepackage{xcolor}
\usepackage{color}
\usepackage{tabularx}

\newcommand\bso{\bgroup\markoverwith{{\rule[0.5ex]{2pt}{0.4pt}}}\ULon}

\usepackage{hyperref} \hypersetup{colorlinks=true,citecolor=blue,linkcolor=blue,urlcolor=black}

\def\CS{{\small CS}}
\def\CV{{\small CV}}
\def\SO{{\small SO}}
\def\MP{{\small MP}}
\def\OMP{{\small OMP}}
\def\LASSO{{\small LASSO}}
\def\SISSO{{\small SISSO}}

\def\SIS{{\small SIS}}
\def\RMSE{{\small RMSE}}

\def\LOO{{\small LOO}}\def\LOOCV{{\small LOOCV}}
\def\LTO{{\small LTO}}
\def\suD{{\substack{\scalebox{0.6}{1D}}}}

\def\snD{{\substack{\scalebox{0.6}{nD}}}}
\def\siD{{\substack{\scalebox{0.6}{iD}}}}
\def\snmD{{\substack{\scalebox{0.6}{(n-1)D}}}}
\def\sRMS{{\substack{\scalebox{0.6}{RMS}}}}
\def\SI{\textcolor{black}{Supplementary Materials}}

\def\EA{E\!A}
\def\IE{I\!E}
\def\CN{C\!N}

\makeatletter \renewcommand\frontmatter@abstractwidth{\dimexpr\textwidth\relax} \makeatother 

\begin{document}


\title{\Large SISSO: a compressed-sensing method for identifying the best low-dimensional descriptor in an immensity of offered candidates}

\author{Runhai Ouyang}
\affiliation{Fritz-Haber-Institut der Max-Planck-Gesellschaft, 14195 Berlin-Dahlem, Germany}
\author{Stefano Curtarolo}
\affiliation{Fritz-Haber-Institut der Max-Planck-Gesellschaft, 14195 Berlin-Dahlem, Germany}
\affiliation{Materials Science, Duke University, Durham, 27708, NC, USA}
\author{Emre Ahmetcik}
\affiliation{Fritz-Haber-Institut der Max-Planck-Gesellschaft, 14195 Berlin-Dahlem, Germany}
\author{Matthias Scheffler}
\affiliation{Fritz-Haber-Institut der Max-Planck-Gesellschaft, 14195 Berlin-Dahlem, Germany}
\author{Luca M. Ghiringhelli}
\email{ghiringhelli@fhi-berlin.mpg.de}
\affiliation{Fritz-Haber-Institut der Max-Planck-Gesellschaft, 14195 Berlin-Dahlem, Germany}

\date{\today}

\begin{abstract}
  \noindent
  The lack of reliable methods for identifying {\it descriptors}
  --- the sets of parameters capturing the underlying mechanisms of a materials property ---
  is one of the key factors hindering efficient materials development.
  Here, we propose a systematic approach for discovering descriptors for materials properties, within the framework of compressed-sensing {based} dimensionality reduction.
  \SISSO\ (\underline{s}ure \underline{i}ndependence \underline{s}creening and \underline{s}parsifying \underline{o}perator) tackles immense and correlated features spaces,
  and converges to the optimal solution from a combination of features relevant to the {materials'} property of interest.
  In addition, \SISSO\ gives stable results also with small training sets.
  The methodology is benchmarked with the {quantitative} prediction of the ground-state enthalpies of octet binary materials (using {\em ab initio} data) and applied
  {to the showcase example of predicting} the metal/insulator classification of binaries (with experimental data). 
  Accurate, predictive models are found in both cases.
  For the metal-insulator classification model, the predictive capability are tested beyond the training data:
  It rediscovers the available pressure-induced insulator$\rightarrow$metal transitions and it allows for the prediction of yet unknown transition candidates, ripe for experimental validation. 
  As a step forward with respect to previous model-identification methods, \SISSO\ can become an effective tool for automatic materials development.
\end{abstract}

\maketitle

\section*{Introduction} \vspace{-2mm}

The materials-genome initiative \cite{MGI_OSTP_new} has fostered high-throughput calculations and experiments.
Correspondingly, computational initiatives (e.g., Refs. \cite{aflow, materialsproject.org, oqmd.org, cmr_repository}), have already tackled many thousands of different systems (see \cite{White2013,Luca2015,Kalidindi2015,Ceder_BigData_SA2016,curtarolo:art112,Wolverton_ML_NPJ2016,curtarolo:art124,Tanaka_ML,curtarolo:art81,MeredigICSD_NMAT2013,Fischer06}).
Much of the data of this field is available in the FAIR Repository and Archive of the NOMAD Center of Excellence\cite{nomad-coe.eu,NOMAD_MRS}. 
On close inspection, one realizes that such data collections are so-far inefficiently exploited, and only a tiny amount of the contained information is actually used. 
Despite the number of possible materials being infinite, the request for specific properties --- e.g.,``a material that is stable, non-toxic, with an optical band gap between 0.8 and 3.2 eV'' --- drastically reduces the set of candidates. This implies that, in terms of functional materials, the structural and chemical space of compounds is sparsely populated.
Identifying these few materials --- known materials as well as materials that have not been created to date --- requires an accurate, predictive approach. 

\begin{figure*}[t!]
  \includegraphics[width=0.78\textwidth]{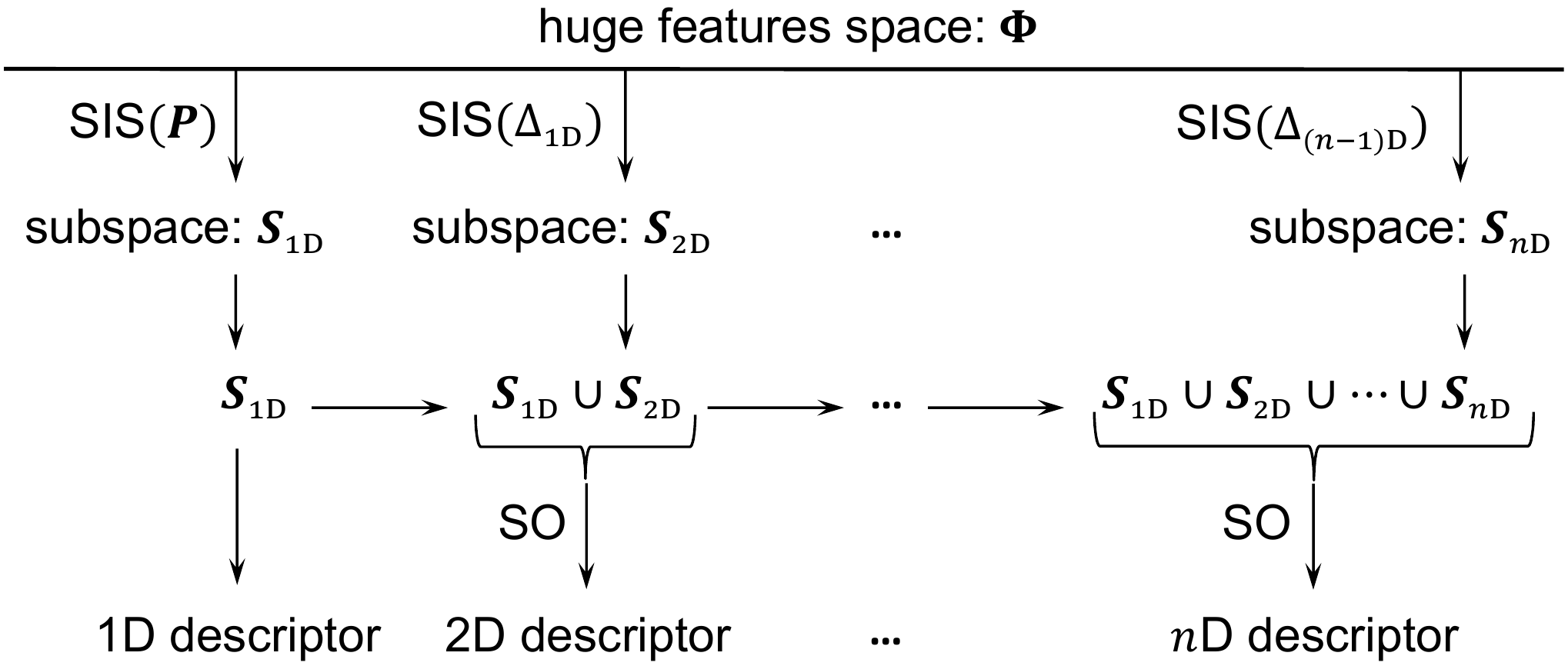}
  \vspace{-1mm}
 \caption{\small
    The method \SISSO\ combines unified subspaces having the largest correlation with residual errors
    $\bm{\Delta}$ (or $\bm{P}$) generated by \SIS\ (sure independence screening) with \SO\ (sparsifying operator)
    to further extract the best descriptor.}
  \label{fig1}
\end{figure*}

Several methods, falling under the umbrella names of artificial intelligence or (big-)data analytics (including data mining, machine/statistical learning, compressed sensing, etc.)
have being developed and applied to the wealth of materials-science data \cite{Bartok_PRL_2010,curtarolo:art85,Rajan_AnnRev2015,Mueller_MLMS_2016_review,Kim_ChemMat_2016,Lilienfeld2016,Takahashi_DalTrans2016,Ceriotti2017,Goldsmith_NJP_2017,Pham_jcp2018}, 
but so far, no general and systematic approach has been established and demonstrated. 
The challenge here is that many different processes and phenomena exist, controlled by atomic structure, electron charge, spin, phonons, polarons and other quasiparticles, and tiny changes in structure or composition can cause a qualitative change of the materials property (phase transitions). 
For example, less than 0.001\% impurities can turn an insulator into a conductor. This type of complexity is a significant element of `the fourth paradigm in materials science' \cite{Hey_Forth_Paradigm_Book_Microsoft,Agrawal_APLM_2016,NOMAD_MRS} which recognizes that it may not be possible to describe many properties of functional materials by a single, physically founded model, i.e., via a closed, analytical expression. The reason is that such properties are determined by several multi-level, intricate theoretical concepts.
Thus, insight is obtained by searching for structure and patterns in the data, which arise from functional relationships (including but not limited to linear correlations) with different processes and functions. Finding a descriptor --- the  set of parameters capturing the underlying mechanism of a given materials property or function --- that reveals these relationships is the key, intelligent step. Once the descriptor has been identified, essentially every learning approach (e.g., regressions --- including kernel-based ones ---, artificial neural networks, etc.) can be applied straightforwardly. 
These issues and in particular the central role of the descriptor was implicitly assumed in many seminal machine-learning works applied to materials science, but it was only later explicitly identified in the works of Ghiringhelli {\em et al.} \cite{Luca2015,Luca2017}.
These authors recast the descriptor-search challenge into a compressed-sensing (\CS) formulation.
The \CS\ approach has been shown to be effective for reproducing a high quality ``reconstructed signal'' starting from a very small set of ``observations'' \cite{CandesWakin_IEEE_SPM_2008,Nelson2013}.
Mathematically, given a set of samples measured incoherently, $\bm{P}$, \CS\ finds the sparse solution $\bm{c}$ of an underdetermined system of linear equations $\bm{D}\bm{c}=\bm{P}$ ($\bm{D}$ is called the {\em sensing matrix} with columns $\gg$ rows).
If the number of nonzero entries in $\bm{c}$ is smaller than the size of $\bm{P}$, then \CS\ effectively reduces the dimensionality of the problem \cite{CandesRombergTao_IEEE_TIT_2006,Donoho2006,CandesWakin_IEEE_SPM_2008}. 
In the specific case treated in \cite{Luca2015,Luca2017}, given a set of materials $m_i$ with observable properties listed in vector $\bm{P}_i$ and a huge list of possible test features $d_j$ (forming the features space), the linear projection of each $i$-material into the $j$-feature forms the $i,j$ components of the sensing matrix $\bm{D}$.
The sparse solution of ``$\operatorname*{arg\,min}_{\bm{c}}\left( \lVert\bm{P}-\bm{Dc}\rVert^2_2 +\lambda\lVert{\bm{c}}\rVert_0 \right)$'', 
where $\lVert{\bm{c}}\rVert_0$ is the number of nonzero components of $\bm{c}$, gives the optimum $n$-dimensional descriptor, i.e., the set of features ``selected'' by the the $n$ non-zero components of the solution vector $\bm{c}$.

In Refs. \cite{Luca2015,Luca2017}, a modification of LASSO (\underline{l}east \underline{a}bsolute \underline{s}hrinkage and \underline{s}election \underline{o}perator, \cite{LASSO}) was introduced for finding the optimal solution.
However, moving beyond the showcase application demonstrated in those papers (predicting the ground-state crystal structure of octet binaries semiconductors), it turns out that the method is unable to deal with large feature spaces, i.e. with situations where knowledge about the underlying processes is not well developed and when in addition to the atomic properties, also collective properties, e.g. the electronic band structure, play a role. When the space of candidate descriptors (the feature space) gets large (larger than few thousands elements) and/or when features are correlated, the approach breaks down. 

In the present paper, we provide a strong and efficient solution of these problems, i.e. we present a new method, called SISSO (\underline{s}ure \underline{i}ndependent \underline{s}creening and \underline{s}parsifying \underline{o}perator), which can deal with an immensity of offered candidate descriptors (billions, or more) and does not suffer when features are correlated. 
The outcome of SISSO is a mathematical model, in the form of explicit, analytic functions of basic, input physical quantities. This aspect gives the opportunity to inspect the equations and suggest means to test the generalization ability of the model.

\section*{Results and Discussion}  \vspace{-2mm}
\noindent {\bf Features e
space construction.}
All quantities that are  hypothesized to be relevant for describing the target property (the so called {\em primary features} \cite{Luca2015,Luca2017}) are used as starting point for the construction of the space \cite{Sondhi2009,Guyon2003}. 
Features are of atomic (species per se) and collective origin (atoms embedded in the environment).
Then, a combination of algebraic/functional operations is recursively performed for extending the space.
For instance, the starting point $\bm {\Phi}_0$ may comprise readily available and relevant properties, such as atomic radii, ionization energies, valences, bond distances and so on.
The operators set is defined as 
\vspace{-1mm}
$$
\bm{\hat{H}}^{\rm (m)} \equiv \left\{I, +,-,\times,/,{\rm exp},{\rm log},|-|,\sqrt{\ },^{-1},^2,^3\right\}\left[\phi_1,\phi_2\right],
\vspace{-1mm}
$$ where
$\phi_1$ and $\phi_2$ are objects in $\Phi$ (for unary operators only $\phi_1$ is considered) and the superscript $^{\rm (m)}$ indicates that
dimensional analysis is performed to retain only ``meaningful'' combinations (e.g., no unphysical items like 
`{\em size$\,+\,$energy}' or `{\em size$\,+\,$size$^2$}').
The intrinsically linear relationship observables $\leftrightarrow$ descriptor in the \CS\ formalism is made non-linear by equipping the features space with non-linear operators 
in $\bm{\hat{H}}^{\rm (m)}$.
At each iteration, $\bm{\hat{H}}^{\rm (m)}$ operates on all available combinations, and the features space grows recursively as:
\vspace{-2mm}
\begin{equation} \label{eq:phin}
  \bm{\Phi}_n \equiv \bigcup_{i=1}^{n} \bm{\hat{H}}^{\rm (m)} \left[\phi_1,\phi_2\right], \quad \forall \phi_1,\phi_2 \in \Phi_{i-1}.
  \vspace{-2mm}
\end{equation}
The number of elements in $\bm{\Phi}_n$ grows very rapidly with $n$.
It is roughly of the order of $\sim (\#\bm{\Phi}_0)^{2^n}\!\times (\#\bm{\hat{H}_2})^{2^n-1}$
where $\#\bm{\Phi}_0$ and $\#\bm{\hat{H}_2}$ are the numbers of elements and binary operators in $\bm{\Phi}_0$ and $\bm{\hat{H}}$, respectively.
For example, $\# \bm{\Phi}_3 \sim 10^{11}$ with $\#\bm{\hat{H}_2}=5$ and $\#\bm{\Phi}_0=10$. 
To avoid {\it a priori} bias and contrary to previous works \cite{Sondhi2009}, no features were disregarded despite the size of the resulting features space.
Instead, we extend the sparse-solution algorithm (using \underline{s}parsifying \underline{o}perators (\SO)~\cite{Breen2009}) 
and tackle huge sensing matrices representative of features spaces containing coherent elements {overcoming the limitations of \LASSO\ based methods \cite{Luca2015,Luca2017}.}

\begin{figure*}[t!]
  \includegraphics[width=0.99\textwidth]{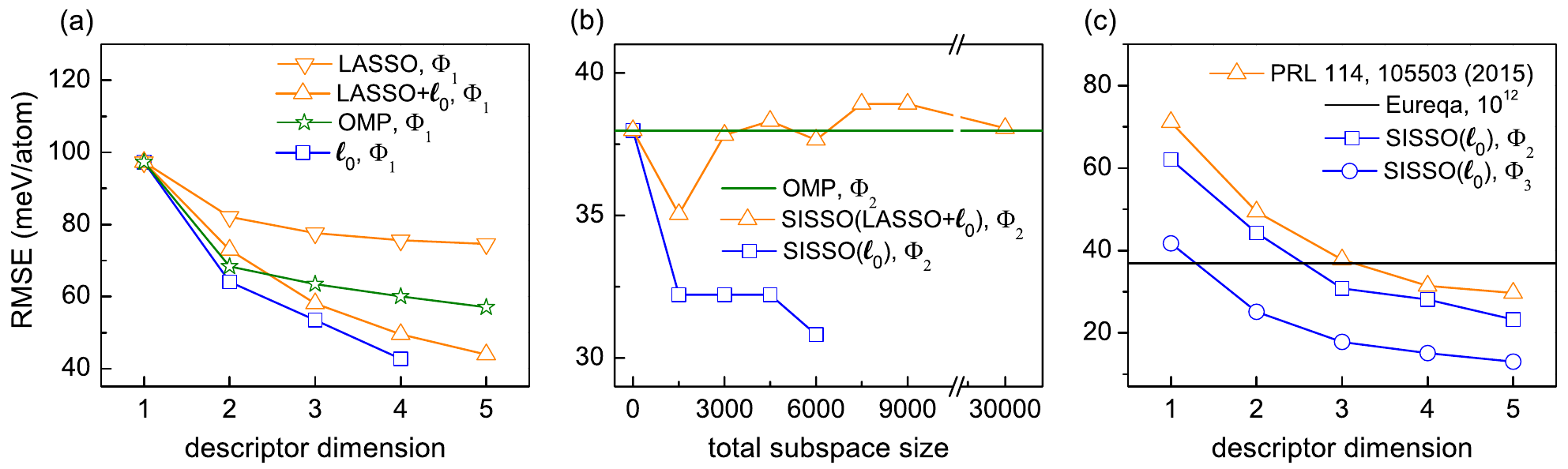}
 \vspace{-1mm}
  \caption{\small
    {\bf Benchmark of algorithms.}
    {\bf (a)} Training error: \RMSE\  {\it versus} descriptor dimension for different \SO s operating on the smallest $\bm{\Phi}_1$.
    {\bf (b)} Training error: \RMSE\ {\it versus} subspace size in the \SIS\ step to find a 3D descriptor by \OMP\ or \SISSO\ with the same large features space $\bm{\Phi}_2$ (a similar picture for a 2D descriptor is presented in the \SI).
    {\bf (c)}  Training error: \RMSE\ by \SISSO($\ell_0$) with $\bm{\Phi}_2$ and $\bm{\Phi}_3$ compared with previous work \cite{Luca2015}
    (features space size $\sim4\,500$) and with the Eureqa software \cite{Schmidt2009} (evaluated functions $10^{12}$, larger than $\#\bm{\Phi}_3$).
   }
  \label{fig2}
\end{figure*} 

\noindent {\bf Solution algorithm.} The $\ell_0$-norm regularized minimization \cite{SISSOnoteELLE0}
is the obvious path for finding the best sparse solution of linear equations.
It is performed through combinatorial optimization by penalizing the number of non-zero coefficients. 
The algorithm is NP-hard and thus infeasible when the {features space becomes very large.}
Efficient methods can be employed to {\it approximate} the correct $\ell_0$ solution~\cite{Donoho2003} 
with ideal features space (e.g., having uncorrelated basis sets).
Amongst them are the convex optimization by $\ell_1$-norm \cite{SISSOnoteELLE1}
regularization  \LASSO\ \cite{LASSO}) and the various greedy algorithms such as the \underline{m}atching \underline{p}ursuit (\MP)~\cite{MP} and \underline{o}rthogonal \underline{m}atching \underline{p}ursuit (\OMP)~\cite{OMP,Tropp2007}.
Unfortunately, with correlated features spaces, approximated results can largely deviate from the ideal $\ell_0$ solutions \cite{Donoho2003,Tropp2004}. 
Corrections have been proposed, for example the \LASSO+$\ell_0$ scheme comprising \LASSO\ prescreening and subsequent $\ell_0$ optimization \cite{Luca2015,Luca2017}, and the
the $\ell_1$-analysis and $\ell_1$-synthesis \cite{l1-analysis}. 
However, when the features space size becomes of the order of $10^6-10^9$, $\ell_1$ based methods also become computationally infeasible. 
As previously mentioned, here we overcome the huge size of the problem by combining \SO\ with \underline{s}ure \underline{i}ndependence \underline{s}creening (\SIS)~\cite{SIS1,SIS2}, 
which has been shown to be effective for dimensionality reduction of ultra-high dimensional features spaces \cite{SIS1}. 
\SIS\ scores each feature (standardized) with a metric (correlation magnitude, i.e., the absolute of inner product between the target property and a feature)
and keeps only the top ranked \cite{SIS1}. 
After the reduction, \SO\ 
is used to pinpoint the optimal $n$-dimensional descriptor. 
The smaller the dimensionality, the better the outcome: progressively larger $n$ are tested until the ``left-over'' residual error is within quality expectation.
The combination of \SIS\ and \SO\ is called \SISSO. Figure \ref{fig1} illustrates the idea.

\noindent {\bf \SISSO.} Out of the huge features space ($\sim10^{10}$ elements or more),
\SIS\ selects the subspace $\bm{S}_\suD$ containing the features having the largest correlation with the response $\bm{P}$ (target material property).
{
  Generally, the larger the subspace $\cup \bm{S}_\siD$, the higher the probability it contains the optimal descriptor.
  However, the chosen size of $\cup \bm{S}_\siD$, depends on {\em i)} which type of \SO\ is later used, {\em ii)} the dimensionality $n$ requested, and {\em iii)} the available computational resources.
  With \SO(\LASSO), $\cup \bm{S}_\siD$ can contain as much as $10^5\sim10^6$ elements, depending on $\#\bm{P}$.
  With \SO($\ell_0$), the largest obtainable size is typically $10^5$ for $n=2$, $10^3$ for $n=3$, $10^2$ for $n=4$, etc. (because the number of needed evaluation grows combinatorially with $n$). 
  If $n$ is large, e.g., $>$10, then the maximum possible $\# \bm{S}_\siD$ converge to 1: \SISSO\ becomes \OMP. }
From inside $\bm{S}_\suD$, \SO($\ell_0$) finds the best 1D descriptor, which is trivially the first ranked feature. In other words, the \SIS\ solution in 1D is already the \SISSO\ solution.
The residual error for a $n$-dimensional model is defined as
$
\bm{\Delta}_\snD \equiv \bm{P}-\bm{d}_\snD\bm{c}_\snD,
$
where $\bm{d}_\snD$ is the matrix with columns being the selected features from the whole features space, and the $\bm{c}_\snD=({\bm{d}_\snD}^T\bm{d}_\snD)^{-1}{\bm{d}_\snD}^T\bm{P}$ is the least square solution of fitting $\bm{d}_\snD$ to $\bm{P}$. 
If the error, the root-mean-square of the residual $\rho_\sRMS(\bm{\Delta}_\snD)$, is below a certain threshold then descriptor is considered fit.
Otherwise the method recursively considers a higher dimensional solution.
In general, for a $n$-dimensional descriptor, \SIS\ selects the subspace $\bm{S}_\snD$ with response $\bm{\Delta}_\snmD$. 
Then \SO\ extracts the best $n$D descriptor, with response $\bm{P}$, from the union of all the previously selected subspaces $\bm{S}_\snD \cup \bm{S}_\snmD \cup \cdots \cup \bm{S}_\suD$.
Cand\`es and Romberg \cite{CandesRomberg_InvProb_2007} have shown that to identify the best $n$-dimensional descriptor with ``overwhelming probability'' 
the size of the response --- in our case the number of materials observations $\bm{P}$ --- 
needs to satisfy the relationship $\#\bm{P} \ge k\cdot n \cdot {\rm log} (\#\Phi)$, where $k$ is a constant (around $1\sim10$ \cite{Luca2017}) and $\#\Phi$ is the size of the features space \cite{CandesWakin_IEEE_SPM_2008}.
Differently from the typical \CS\ scenario, here $\#\bm{P}$ is fixed \cite{Luca2017}; then, when $\#\Phi$ increases, the maximum $n$ decreases in order to satisfy the relationship \cite{CandesRomberg_InvProb_2007}.
In practice, features spaces of growing sizes ($\Phi_0, \Phi_i, \cdots$) and different $n$ are tested until a model with required accuracy ($\rho_\sRMS(\bm{\Delta}_\snD) <$ threshold) is obtained. 

\SISSO\ has advantages over \MP\ \cite{MP} and \OMP\ \cite{OMP} .
\MP\ searches a linear model reproducing $\bm{P}$ by adding dimensionality to a descriptor while preserving selected features and corresponding coefficients.
\OMP\ improves \MP\ by reoptimizing the coefficients every time a new component is introduced, $n\rightarrow{n+1}$, but still preserving previously selected features. 
\SISSO\ both reselects features and reoptimizes coefficients at each dimensional increment.
\SISSO\ reduces to \OMP\ when each subspace in the union has unit size ($\#\bm{S}_\siD=1, \forall i$). Still, it differs from iterative \SIS\ \cite{SIS1} which reduces to simple \MP\ when all $\#\bm{S}_\siD=1$. 

\noindent {\bf Benchmark: Quantitative prediction.}
\SISSO\ is benchmarked by comparing the relative stability of octet binary materials between rock-salt (RS) and zinc-blende (ZB) configurations.
The reference data is taken from Ref. \cite{Luca2015}, including the target calculated {\it ab initio} enthalpy difference, RS and ZB for 82 materials and the 23 primary features related to material compositions forming $\Phi_0$.
The primary features {considered in this study} are listed in the \SI. All quantities are calculated with density-functional theory in the local-density approximation. Details are given in Refs. \cite{Luca2015,Luca2017}.
Then, with a combination of the previously defined operator set, $\bm{\hat{H}}^{\rm (m)}$, and Eq. (\ref{eq:phin}), the features spaces $\bm{\Phi}_1$ (small, $\#\bm{\Phi}_1=556$), $\bm{\Phi}_2$ (large, $\#\bm{\Phi}_2\!\sim\!10^5$), and $\bm{\Phi}_3$ (huge, $\#\bm{\Phi}_3\!\sim10^{11}$) are constructed.

\noindent {\bf Figure \ref{fig2}(a).}
The training errors ($\rho_\sRMS$) of different \SO: \LASSO, \LASSO+$\ell_0$, \OMP, and $\ell_0$ are compared while operating on the small features space $\bm{\Phi}_1$.
\LASSO\ suffers because of the correlations existing inside $\bm{\Phi}_1$ (Figure S1 in the \SI);
\LASSO+$\ell_0$ and \OMP\ both surpass \LASSO;
$\ell_0$ is the reference: it gives the exact global minimum solution for descriptors of any dimension.
However, even with $\ell_0$ the error is {still too large} for {many} thermodynamical predictions --- $\rho_\sRMS(\bm{\Delta}_\snD) \gtrsim 40$ meV/atom --- and this is due to the too-small size of $\bm{\Phi}_1$. \\
\noindent {\bf Figure \ref{fig2}(b).}
For the larger $\bm{\Phi}_2$, \SIS\ combined with \LASSO+$\ell_0$ as \SO\ --- \SISSO(\LASSO+$\ell_0$) ---, \SISSO($\ell_0$), and \OMP\ are compared for generating a 3D descriptor:
\SISSO($\ell_0$) is the only approach improving consistently with subspace size $\#\!\cup\!\bm{S}_\siD$ and it always surpasses \OMP\ when each $\#\!\bm{S}_\siD\!\gg\!1$;
\SISSO(\LASSO+$\ell_0$) does not improve over \OMP\ because of the failure of \LASSO\ in dealing with correlated features \cite{Donoho2003}.
Obviously, the larger the features space and the better the obtainable model (at least equal). 
When exhaustive searches become computationally impossible, \SISSO\ can still find the optimal solution if the subspace produced by \SIS\ is big enough.\\
\noindent {\bf Figure \ref{fig2}(c).}
The errors for 1- to 5-dimensional descriptors are calculated by \SISSO($\ell_0$) while operating in the large $\bm{\Phi}_2$ and huge $\bm{\Phi}_3$ spaces. 
For $n=1$, \SIS\ reduces to the best 1D descriptor, so no $\ell_0$ is needed. For $n=2,3,4,5$ the size of the \SIS\ subspace 
is chosen to follow the previously mentioned relationship \cite{CandesRomberg_InvProb_2007} applied to the subspace $\#\bm{S} \sim {\rm exp} (\#\bm{P}/kn)$.
With $\#\bm{P}=82$ and $k=3.125$, the total size of all the selected subspaces is $\#\cup\bm{S}_\siD= 5\cdot10^5, 6\cdot10^3, 7\cdot10^2, 2\cdot10^2$ for $n=2,3,4,5$, respectively.
For all these sizes, the application of $\ell_0$ regularization as \SO\ involves $10^{10}$--$10^{11}$ independent least-square-regression evaluations. This is computationally feasible due to our (trivially) parallel implementation of \SISSO\ (for instance, for this application, the production calculations were run on 64 cores).
The training errors for the descriptors identified from $\bm{\Phi}_3$ are systematically better than those coming from $\bm{\Phi}_2$, thanks to the higher complexity (functional forms of the descriptors are reported in \SI).
\SISSO($\ell_0$) with $\bm{\Phi}_2$ is systematically better than the previous work by Ghiringhelli {\it et al.} \cite{Luca2015,Luca2017}, due to the allowed larger features spaces.
Note that when \SISSO($\ell_0$) is applied to the same features space as in Ref. \cite{Luca2015}, it also finds the same descriptor: \SISSO\ combined with the features space of Ref. \onlinecite{Luca2015} has the same results of the yellow line of Figure \ref{fig2}(c). 
Performance is also compared with the commercial software Eureqa~\cite{Schmidt2009} by using the same operator set and primary features ($\bm{\Phi}_0$), and $10^{12}$ evaluated functions, a number comparable to $\#\bm{\Phi}_3$.
\SISSO($\ell_0$) in $\bm{\Phi}_3$ with $n\ge2$ and \SISSO($\ell_0$) in $\bm{\Phi}_2$ with $n\ge3$ have both lower training error than Eureqa (see \SI). \\

\begin{figure*}
  \includegraphics[width=0.73\textwidth]{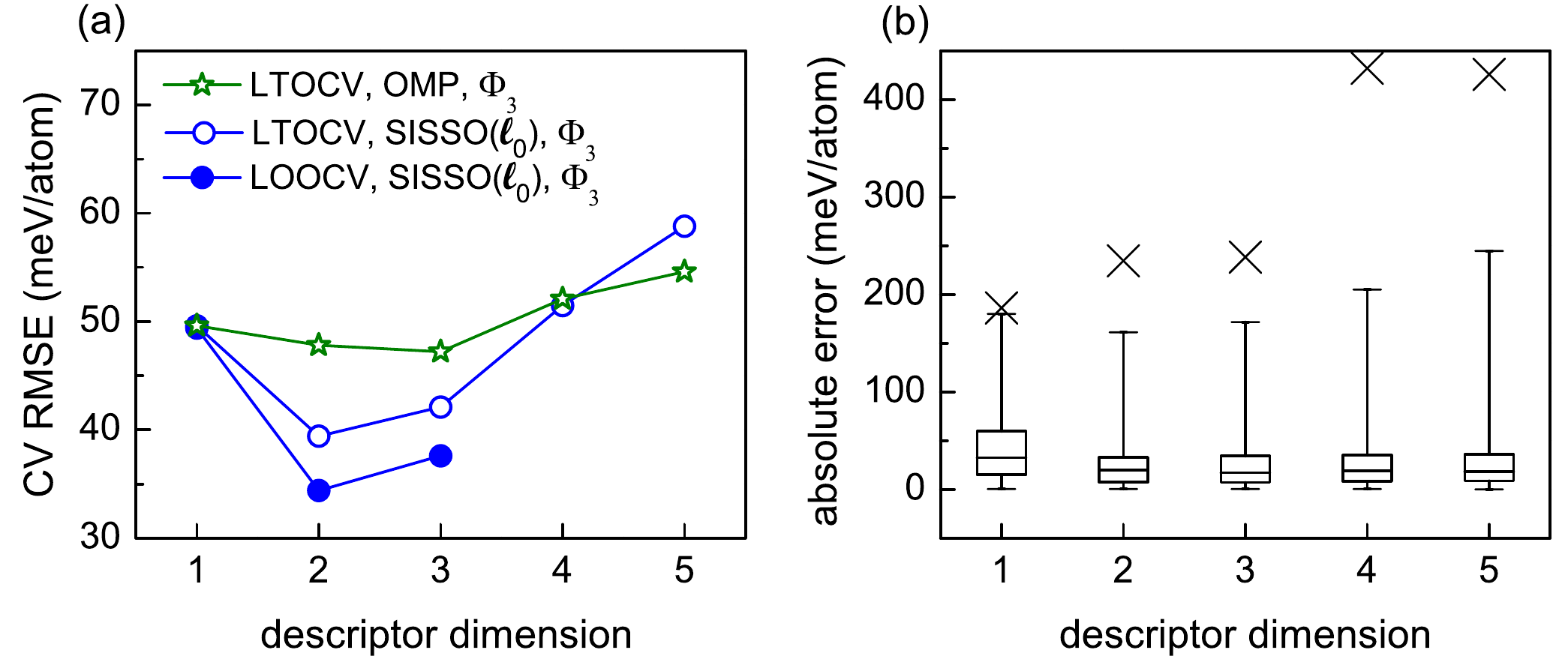}
  \vspace{-2mm}
  \caption{\small
    {\bf Benchmark of algorithms.}
    {\bf (a)} Cross validation: \LTO\CV\ and \LOOCV\ results for the features space $\Phi_3$ with \OMP\ and \SISSO($\ell_0$).
    {\bf (b)} Cross validation: Box plots of the absolute errors for the \SISSO($\ell_0$)-\LTO\CV\ results with features space $\Phi_3$. 
      The upper and lower limits of the rectangles mark the 75\% and 25\% percentiles of the distribution,
      the internal horizontal line indicates the median (50\% percentile), 
      and the upper and lower limits of the ``error bars'' depict the 99\% and 1\% percentiles. 
      The crosses represent the maximum absolute errors.
  }
  \label{fig3}
\end{figure*} 

\noindent {\bf Figure \ref{fig3}(a).}
Training errors were illustrated in Figure \ref{fig2}(a-c), in order to directly compare over the same dataset, the ability of different approaches to find optimal or close-to-optimal solutions of the \CS\ problem. 
With practical applications in mind, it is imperative to determine the performance of the obtained model on data that are not used for the training.
In statistical learning \cite{Hastie2008,Pearl_Book_Inference}, this is performed via {\it cross validation} (\CV), a class of techniques that, by splitting the dataset into a training and a test set in various ways, aims at detecting ``underfitting'' and ``overfitting'', 
i.e., when the complexity of the fitted model is too small or too large, respectively. In \CS, dedicated \CV\ techniques have been proposed \cite{Boufounos_CompressSensing_2007,Ward_CompressSensing_2008}.
Specifically, in a \CS-based iterative technique like \SISSO, the only source of overfitting can come from a too large dimensionality of the descriptor (note that there is only one fitting coefficient per dimension{, i.e., features recursively built via Eq. \ref{eq:phin} do not contain fitting parameters}).
For this benchmark application, we applied the \CS-\CV\ scheme proposed in Ref. \cite{Boufounos_CompressSensing_2007} with leave-10\%-out (\LTO) \CV\ 
(the dataset is split in 40 training set containing 90\% randomly selected data points and a test set with the remaining 10\%)
and leave-one-out (\LOO) \CV\ (one data points constitutes the test set, and the procedure is iterated $\#P$ times). 
The model is trained on the training set (the whole \SISSO\ procedure, i.e., including the selection of the descriptor) and the error is measured on the test set. 
{In such framework, the \CV\ error decreases with the number of iterations --- the dimensionality --- 
  until the approximate descriptor will try to fit the data (containing possible errors) starting from primary features having intrinsic limitations, thus causing a subsequent increase in the \CV\ error. }
The iteration at which the \CV\ error starts increasing identifies the maximum dimensionality of that particular model. This is determined by the features space --- in turns determined by set of primary features, operators set, and number of iterations of the features space construction --- and the training set.
{\CS-\CV\ is performed for $\Phi_3$ with the subspace sizes reported in the description of Figure \ref{fig2}(c), and for subspace of unit size (for which \SISSO\ becomes \OMP).
  It is found that the dimensionality minimizing the error is two for both the \CV\ schemes of \SISSO($\ell_0$). In order to achieve a smaller prediction error, one would then need to add new primary features, possibly substituting features that are never selected in a descriptor, or increase the complexity of the features space, or both.
  \OMP\ finds the same dimensionality of the problem (2$\sim$3), has a lower computational cost but a cost of worse performance in terms of prediction error.} \\
\noindent {\bf Figure \ref{fig3}(b)} depicts the box plots for the distribution of errors as function of the dimensionality for \SISSO($\ell_0$)-\LTO\CV\ results with features space $\Phi_3$ (\RMSE\ shown in (a)). 
  The 1\% and 99\% (extrema of the ``error bar''), the 25\% and 75\% (lower and upper limits of the rectangle) and the median (intermediate horizontal line) percentiles are marked.
  The maximum absolute errors are also indicated by crosses.
  The worsening of the \RMSE\ beyond 2D is mainly determined by an increase in the largest errors (the 99\%-percentile), while most of the errors remain small (median/lower percentiles $\sim$ constant). \\
\noindent{\LOOCV\ is also used to inspected how often the same descriptor is selected.} The test operates in $\#\Phi_3$ with \SISSO($\ell_0$). 
The \LOOCV\ descriptor agrees with the one found over all data 79, 73, 58 times out of 82 iterations. 
It is remarkable, as the size of $\Phi_3$ is of the order $10^{11}$ features and there are only 82 data points. 
This means that the 1D, 2D, 3D descriptor is selected from $10^{11}$, $10^{22}$, $10^{32}$ 
combinations, respectively. 
We note that descriptors that are selected using the reduced training data set need be correlated with the full data-set descriptors, implying the existence of a ``hidden'' correlation between the functional forms. 
Hence, selecting different descriptors does not imply over-fitting (this is independently determined via \CS-\CV), but choosing different existing approximate functional relationship among the primary features. 

\begin{figure*}[t]
  \includegraphics[width=0.96\textwidth]{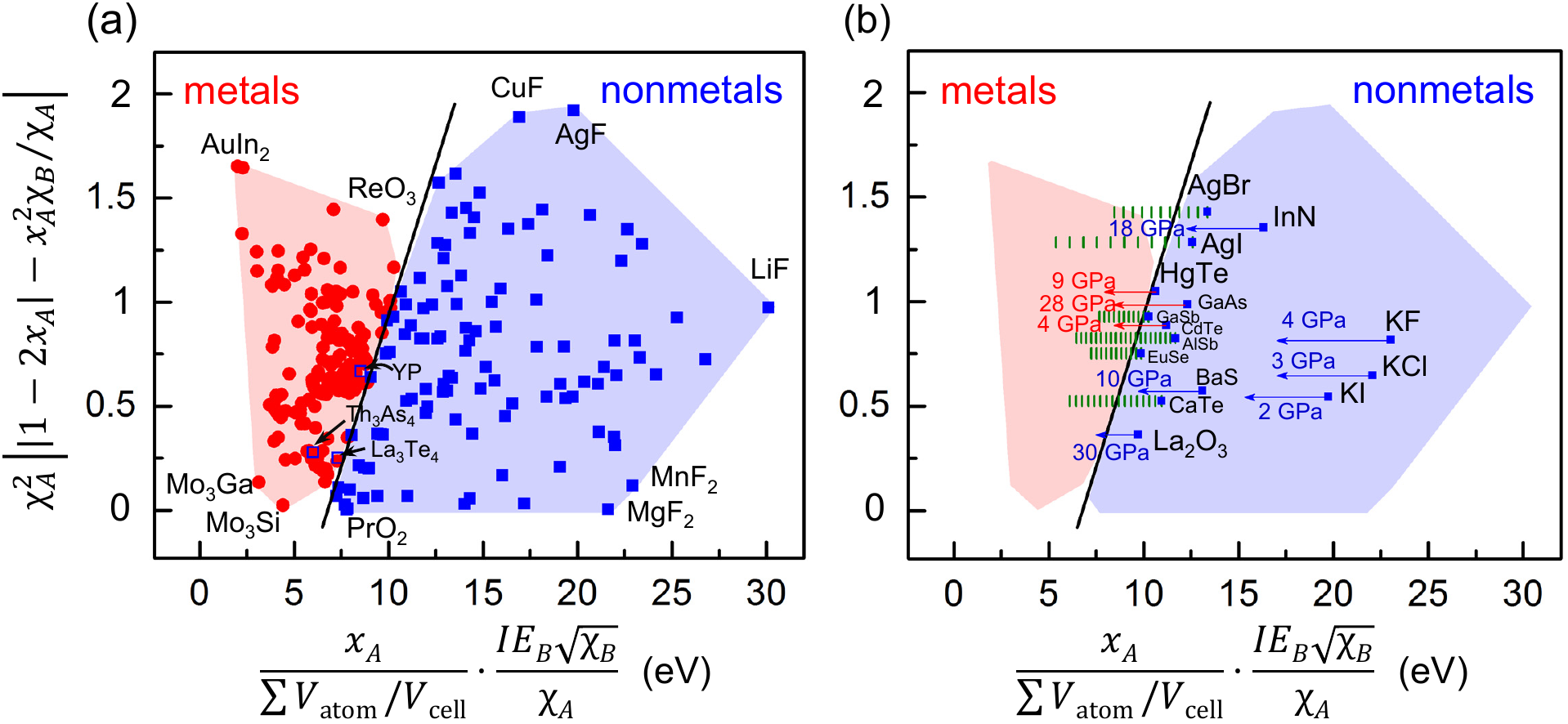}
  \vspace{-2mm}
  \caption{\small
    {\bf \SISSO\ for classification.}
    {\bf (a)}
    An almost perfect classification (99.0\%) of metal/nonmetal for 299 materials.
    Symbols: $\chi$ Pauling electronegativity, $I\!E$ ionization energy,
    $x$ atomic composition, $\sum V_{\rm atom}/V_{\rm cell}$ packing factor. 
    Red circles, blue squares, and open blue squares represent metals, non-metals, and the three erroneously characterized non-metals, respectively.
    {\bf (c)} 
    Reproduction of pressure induced insulator$\rightarrow$metals transitions (red arrows), of materials that remain insulators upon compression (blue arrows), and computational predictions at step of 1GPa (green bars). 
  }
  \label{fig4}
\end{figure*}

\noindent {\bf Application: classification models.}
The \SISSO\ framework can be readily adapted to predict categorical properties (as opposed to continuous properties like an energy difference), i.e., it can be applied for classification.
In the space of descriptors, each category's domain is approximated as the region of space (area, in 2D) within the convex hull of the corresponding training data.
\SISSO\ finds the low-dimensional descriptor yielding the minimum overlap (or maximum separation) between convex regions.
Formally, given a property with $M$ categories, the norm for classification is defined as:
\begin{equation}\label{eq:L0}
  \bm{\hat{c}} \equiv \operatorname*{arg\,min}_{\bm{c}}\left(\sum_{i=1}^{M-1}\sum_{j=i+1}^{M}O_{ij}+\lambda\lVert{\bm{c}}\rVert_0 \right),
\end{equation}
where $O_{ij}$ is the number of data in the overlap-region between the $i-$ and $j-$domain,
$\bm{c}$ is a sparse vector (0/1 elements) so that a feature $k$ is selected(deselected) when $c_k=1(0)$,
and $\lambda$ is a parameter controlling the number of nonzero elements in $\bm{c}$. 
Of all the possible solutions of Eq.~(\ref{eq:L0}) having the same dimension and overlap,
we chose the one with minimum $n$-dimensional overlap volume \cite{Bialon2016}:
\begin{equation}
  \Omega\equiv\frac{2}{M(M-1)}\sum_{i=1}^{M-1}\sum_{j=i+1}^{M}\frac{\Omega_{ij}}{\min(\Omega_i,\Omega_j)},
  \label{eq:overlapS}
\end{equation}
where $\Omega_{i}$, $\Omega_{j}$, and $\Omega_{ij}$ are the $n$-dimensional volumes of the $i-$, $j-$, and overlap $ij-$domains. 
Finally, the \SIS\ correlation ``property$\leftrightarrow$feature'' is defined as $\left(\sum_{i=1}^{M-1}\sum_{j=i+1}^{M}O_{ij}+1\right)^{-1}$: high correlation $\Leftrightarrow$ low overlap.

\noindent \SISSO\ for classification is tested on a simple metal/nonmetal classification of binary systems. 
The training systems are far from creating an exhaustive list and, as such, the test is strictly meant for benchmarking the validity and implementation of Equations (\ref{eq:L0}-\ref{eq:overlapS}). 
All essential atomic and structural parameters are included as primary features in $\bm{\Phi}_0$.
They originate from the WebElements \cite{webelements} (atomic) and SpringerMaterials \cite{springer} (structural) databases and they are listed in the \SI.
Amongst them are the Pauling electronegativity $\chi$, ionization energy $I\!E$, covalent radius $r_{\rm cov}$, electron affinity, valence (number of valence electrons for $A$ and (8-valence) for $B$), coordination number, interatomic distance between $A$ and $B$ in crystal, atomic composition $x_A$, and  a ``packing parameter'', here the normalized ratio between the volume of spherical atoms and the unit cell: ${\sum V_{\rm atom}/V_{\rm cell}}$ with $ V_{\rm atom}=4\pi r_{\rm cov}^3/3$.
The operator set $\bm{\hat{H}}^{\rm (m)}$ and Eq. (\ref{eq:phin}) are then used to generate $\bm{\Phi}_3$ ($\sim10^{8}$ elements).
Note that \SISSO\ finds its optimal descriptor based on combinations of the input physical quantities (features): non-optimal outcomes indicate that the target property depends on features not yet-considered in $\bm{\Phi}_0$.
As such, to avoid ``garbage in, garbage out'', \SISSO\ requires physical intuition in the choice of features to add:
conveniently, important and non-important features will be automatically promoted or neglected.
Here, since metallicity also depends on ``interstitial charge'', the inclusion of a packing parameter related to superpositions of orbitals is advantageous. 
Given a set of features, \SISSO\ finds their best combination leading to the optimum descriptor. 
If the packing parameter were removed from the primary list, \SISSO\ would autonomously select the combination of features trying to replicate as much as possible the lost descriptive power, in this case the $AB$ atomic distances \cite{OuyangJCP07}.
The experimental binary data set, extracted from the SpringerMaterials database \cite{springer} and used for training the \SISSO\ model, contains $A_xB_{1-x}$ materials having:
{\bf i.} every possible $A$ species;
{\bf ii.} $B$ as $p-$block element (plus H and with the condition $A\neq B$, i.e., elemental solids, such as carbon diamond, are not tackled);
{\bf iii.} non-layered structure and without dimers (the coordination polyhedron of $A$ comprises only $B$ atoms, and vice versa);
{\bf iv.} good experimental characterization and without large distortions (we do not have any distortion feature).
A total 299 binaries in 15 prototypes (NaCl, CsCl, ZnS, CaF$_2$, Cr$_3$Si, SiC, TiO$_2$, ZnO, FeAs, NiAs, Al$_2$O$_3$, La$_2$O$_3$, Th$_3$P$_4$, ReO$_3$, ThH$_2$) are then used. The training materials are listed in the \SI.
Details on the feature-space construction and model identification are given in Appendix.
Out of $\bm{\Phi}_3$, \SISSO($\ell_0$) identifies a 2-dimensional
descriptor with a training accuracy of $\sim99.0$\%. 
The convex domains, indicating metallic and non-metallic materials, are shown in Figure \ref{fig4}.
The figure also includes a line calculated with a support-vector machine \cite{SVM1998}, to help visualizing the separation between convex domains.
These plots are called {\em material-properties maps} (or {\em charts} \cite{Ashby_MAPS_1972,pettifor:1984,pettifor:1986,Luca2015,curtarolo:art94}) 
and \SISSO\ has been specifically designed to identify low-dimensional
regions, {possibly non overlapping}. \\

\noindent {\bf Figure \ref{fig4}(a)} shows the three incorrectly classified systems (blue empty squares).
YP (NaCl prototype)
might have slightly erroneous position in the figure:
the covalent radius $r_{\rm cov}$(Y) (controlling the packing parameter) suffers of large intrinsic errors (see Figure 2 of Ref. \cite{Cordero2008}) and therefore
the compound position might be misrepresented.
La$_3$Te$_4$ and Th$_3$As$_4$ (Th$_3$P$_4$ prototype) are different.
In this case, SISSO indicates that the primary feature are not enough or that the compounds have been experimentally misclassified (due to defects or impurities \cite{La3Te4,Th3As4}).
Inspection of the found descriptor suggests a justification of the involved primary features. The $x$-projection --- $x$-axis in Figure \ref{fig4}(a) ---
indicates that the higher the packing factor $\sum{V_{\text{atom}}/V_{\text{cell}}}$, i.e., the higher the interstitial charge,
the higher the propensity of a material to be a metal. This is not surprising. The merit of the descriptor found by \SISSO\ is to 
$i$) provide a {\em quantitative} account of the dependence of metallicity on the packing factor, allowing for {\em predictions} (see below) and
$ii$) reveal the functional form packing factor $\rightarrow$ metallicity: It is not trivial that the descriptor is linear with the inverse packing factor.
Metallicity also correlates with the electronegativity of the $A$ species, often the main electron donor, 
by competing against the $B$ species, a $p$-element trying to complete its covalent/ionic bonds by filling the unoccupied orbitals and thus removing interstitial charge.
Thus it is not surprising that the material with largest $x$-projection is LiF, a purely ionic compound with closed electron-shells: 
the ratio amongst the two extreme electronegativities, (Li has the lowest, F the highest), pushes the compound toward the rightmost corner of the non-metals domain.
On the other side, AuIn$_2$ is the compound furthest from the non-metals region:
Au has the highest $\chi$ amongst transition metals and In has one of the smallest $\chi$ of the considered $p$-elements.
Available experimental band gaps were also extracted and a figure showing their distribution on the right hand side of the panel is reported in the \SI.
The robustness of the descriptor is corroborated by leave-one-out cross validation.
In 97.6\% of the times, \LOOCV\ reproduces the same functional solution obtained from the whole data. 
In the few cases where the descriptor differs from the all-data one,
the packing fraction always remains; even more: the packing fraction
is present in all features selected by \SIS\ at the first
iteration. 

\noindent {\bf Beyond the training: Prediction of metalization by compression.} 
Although pressure is {neither} included in the features space {nor in the training data}, its effect can be tested by reducing $V_{\rm cell}$.
Amongst the training data, we have 3 systems experiencing pressure-induced insulator$\rightarrow$metal transition: HgTe, GaAs and CdTe.
HgTe, CdTe and GaAs go from insulating zinc blende to metallic rock salt (or an orthorhombic oI4 phase for GaAs) at $\sim$ 9, 4, and 28 GPa, respectively (see red arrows).
Geometrical parameters (cell volumes) at normal and high pressure are taken from the experimental databases and used to modify the $x$-coordinate of the descriptor.
Concurrently, we have also looked for materials that do not become metallic with high-pressure structural transitions (indicated by the blue arrows). In this case our model again makes a correct prediction. 
{\bf Figure \ref{fig4}(b)} shows that the descriptor is perfectly capable of reproducing the correct metallic state.
The idea can be extended to systems which have not yet been fully characterized to predict potential insulator$\rightarrow$metal transitions.
The subset of prototypes which are reasonably close to the domain convex hull and have a fully characterized {\it ab initio} elastic tensors \cite{curtarolo:art115} 
are ``computationally compressed'' by having their $V_{\rm cell}$ reduced following the first order linearized bulk modulus relation: $(V_{\rm cell}(p)-V_{\rm cell}(0))/V_{\rm cell}(0)\sim -p/B_{T}$,
where $p$ is the pressure and $B_{T}$ is the isothermal bulk modulus extracted from the entries in the {\sf AFLOW.org} repository \cite{curtarolo:art115} (see SI for the entries data).
The panel shows a set of compounds for which the descriptor predicts the transition to metallic. 
The green marks are positioned at 1~GPa steps to allow an informed guess of the pressure.
Within this approximation, some compounds are predicted to become metallic at pressure between 5 and 15~GPa: AgBr, AgI, GaSb, AlSb, EuSe, and CaTe.
Pressure-induced structural phase transitions are also not considered in such analysis and thus,
the insulator$\rightarrow$metal transition pressure might be overestimated facilitating experimental validations. \\

\begin{figure}[t]
  \includegraphics[width=0.35\textwidth]{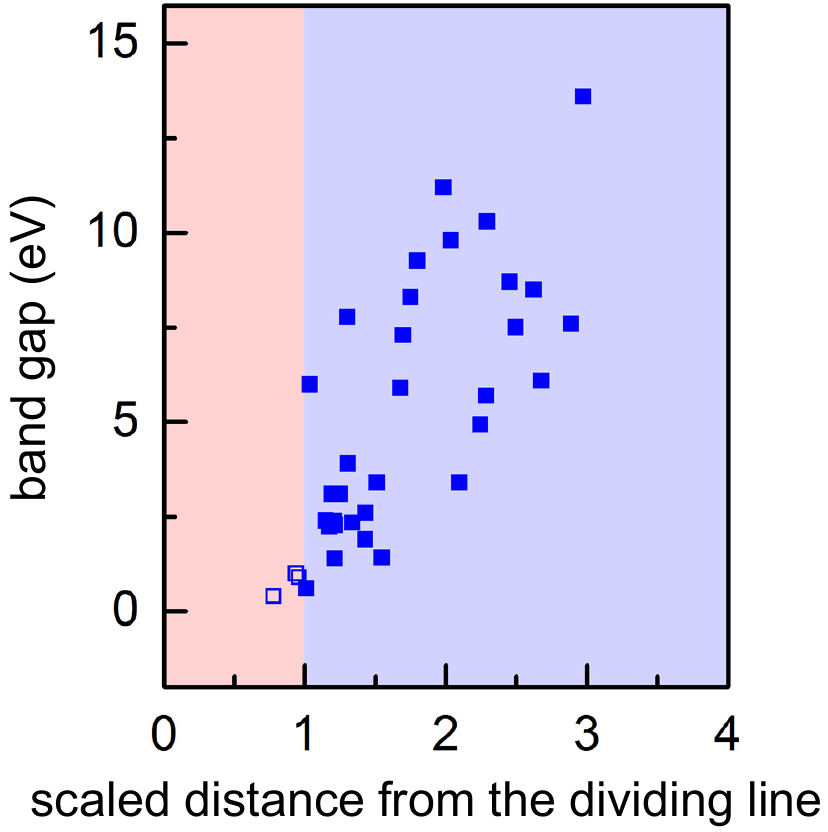}
   \vspace{-1mm}
 \caption{\small
    {\bf \SISSO\ for classification.} 
    Correlation between the band gap of the non-metals and the scaled coordinate from the dividing line.
  }
  \label{fig5}
\end{figure}
\noindent {\bf Beyond the training: Significance of the ``distance'' from the dividing line.}
Figure \ref{fig5} depicts the experimental band gap of the insulators vs. the {\it scaled distance} from the dividing line, i.e., the dimensionless ratio between the $x$-projection of its descriptor versus the $x$-projection of the dividing line corresponding to the $y$-projection of its descriptor value.
With this rescaling, the dividing line corresponds to the vertical line $x=1$
The trend of the data points reveals that the descriptor found by \SISSO\ --- trained only on a categorical property --- 
includes a quantitative, albeit approximate, account of how strongly an insulator is far from being a metal, by locating materials with large band gaps further from the line than small-gap materials.\\

\noindent {\bf General remarks on the descriptor$\rightarrow$property relationship identified by \SISSO}.
As clear from the two application cases presented here, the equations found by SISSO are not necessarily unique and all components of the descriptors may change at each added dimension.
This reflects the approximate nature of the equations and the unavoidable relationships among features (one or more primary features may be accurately described by nonlinear functions of a subset of the remaining features).
We also note that the mathematical constraints imposed in order to obtain solutions efficiently (linear combination of nonlinear functions for the continuous-property case and minimally overlapping convex hulls in the classification case), are very flexible but not complete. I.e., the found descriptor$\rightarrow$property relationship is intrinsically approximate.

\section*{Conclusions}   \vspace{-2mm}
We have presented an efficient approach for extracting effective materials descriptors out of huge and possibly strongly correlated features spaces.
This algorithm, called \SISSO\ (\underline{s}ure \underline{i}ndependence \underline{s}creening and \underline{s}parsifying \underline{o}perators) 
tackles huge spaces while retaining the effectiveness of compressed sensing. 
Specifically, \SISSO\ is built to work also (but not limited to) when only relatively small training sets are available.
\SISSO\ autonomously finds the best descriptor from a combination of features (physical properties),
and it is capable of determining the ones not relevant to the problem, so that the features space can be further optimized.
\SISSO\ identifies the descriptor $\rightarrow$ property relationship in terms of an analytical equation. It does not need to be exact --- a simple, analytical descriptor $\rightarrow$ property function may not even exist --- but it is the most accurate expression given the available features space. If an exact, analytic expression does indeed exist, \SISSO\ is expected to find it if included in the features space.

\SISSO\ shows superior advantages with respect to other established methods, e.g., \OMP\ and \LASSO\ as well as the software Eureqa, especially when dealing with a correlated features spaces.
\SISSO\ does not have the limitation of \LASSO, which suffers with large and highly correlated features spaces. Currently, the only issue of \SISSO\ is the required computer memory needed to handle the features space, and efforts are underway for more efficient implementations.
Our approach is benchmarked on the quantitative modeling of enthalpy differences for a set of zinc-blende and rock-salt prototypes and applied to the metal/insulator classification of binaries.
The robustness of the classification is corroborated by the proper reproduced insulator$\leftrightarrow$metal transitions, which allows to predict a set of systems for further experimental analyses. 

\section*{Acknowledgments}    \vspace{-2mm}
The authors thank Daria M. Tomecka, Cormac Toher, and Corey Oses for their valuable help in collecting the data for the metal/insulator application.
This project has received funding from the European Union’s Horizon 2020 research and innovation program {(\#676580: The NOMAD Laboratory --- an European Center of Excellence and \#740233: TEC1p)}, and the Berlin Big-Data Center (BBDC, \#01IS14013E).
S.C. acknowledges DOD-ONR (N00014-13-1-0635, N00014-11-1-0136, N00014-15-1-2863) and 
the Alexander von Humboldt Foundation for financial support.

\section*{Appendix} 
In this appendix, we present details on the metal/insulator-classification application. \\
\noindent {\bf Primary features}.
Descriptors are to be identified by \SISSO\ from a systematically-constructed large/huge features space in which components are
generated by recursively transforming a set of input primary features, $\Phi_0$, via algebraic operations,
$\bm{\hat{H}}\equiv \{I, +,-,\times,/,{\rm exp},{\rm log},|-|,\sqrt{\ },^{-1},^2,^3\}$.
Primary features usually comprise of properties of isolated atoms (atomic features) and properties of the materials (composition and geometry). 
For the test on binaries' metal/nonmetal classification, the following is the full list of considered primary features:. 
(1) first ionization energy, $\IE_A$ ($A$-species) and $\IE_B$ ($B$-species);
(2) electron affinity, $\EA_A$ and $\EA_B$; 
(3) atom covalent radius, $r_{\rm covA}$ and $r_{\rm covB}$; 
(4) Pauling electronegativity, $\chi_A$ and $\chi_B$; 
(5) valence, $v_A$ (\#valence electrons) and $v_B$ ($8-$\#valence electrons);
(6) coordination number, $\CN_A$ (\#nearest neighbor $B$ of $A$) and $\CN_B$;
(7) interatomic distance between $A$ and $B$ in crystal, $d_{AB}$; 
(8) atomic composition $x_A$ (or $x_B = 1-x_A$; and 
(9) the ratio of the cell volume to the total atom volume in the unit cell of the crystal, $V_{\rm cell}/\sum{V_{\rm atom}}$ ($V_{\rm atom}=4{\pi}r_{\rm cov}^3/3$). 

\begin{table*}[ht]
 \caption{\small \label{tb:TABLE1} Dependence of the metal-insulator classification descriptors on the prototypes of training binary materials.}
\begin{tabularx}{\textwidth}{||X|c|p{3cm}|p{5cm}|c||}
\hline
 prototypes & \#materials & primary features & descriptor & class. \\
\hline
\hline
NaCl & 132 & $\IE_A$, $\IE_B$, $\chi_A$, $\chi_B$, $r_{\rm covA}$, $r_{\rm covB}$, $\EA_A$, $\EA_B$, $v_A$, $v_B$, $d_{AB}$ & $d_1:=\frac{\IE_A\IE_B(d_{AB}-r_{\rm covA})}{exp(\chi_A)\sqrt{r_{\rm covB}}}$ & 100\% \\ 
\hline 
NaCl, CsCl, ZnS, CaF$_2$, Cr$_3$Si & 217 & $\IE_A$, $\IE_B$, $\chi_A$, $\chi_B$, $r_{\rm covA}$, $r_{\rm covB}$, $d_{AB}$, $\CN_A$, $\CN_B$ & $d_1:=\frac{\IE_Bd_{AB}^2}{\chi_Ar_{\rm covA}^2\sqrt{\CN_B}}$, $d_2:=\frac{\IE_A^2r_{\rm covB}\log{(\IE_A)}|r_{\rm covA}-r_{\rm covB}|}{\CN_B}$ & 100\% \\
\hline
NaCl, CsCl, ZnS, CaF$_2$, Cr$_3$Si, SiC, TiO$_2$, ZnO, FeAs, NiAs & 260 & $\IE_A$, $\IE_B$, $\chi_A$, $\chi_B$, $r_{\rm covA}$, $r_{\rm covB}$, $d_{AB}$, $\CN_A$, $\CN_B$ & $d_1:=\frac{d_{AB}/r_{\rm covA}-\chi_A/\chi_B}{\exp{(\CN_B/\IE_B)}}$, $d_2:=\frac{r_{\rm covA}^3d_{AB}\IE_B}{\left|\chi_B/\chi_A-|\CN_B-\CN_A|\right|}$ & 99.6\%\footnote{\label{fn:YP}One entry misclassified: YP-compound in NaCl-prototype.} \\ 
\hline
NaCl, CsCl, ZnS, CaF$_2$, Cr$_3$Si, SiC, TiO$_2$, ZnO, FeAs, NiAs & 260 & $\IE_A$, $\IE_B$, $\chi_A$, $\chi_B$, $x_A$, $x_B$, $V_{\rm cell}/\sum{V_{\rm atom}}$ & $d_1:=\frac{V_{\rm cell}}{\sum{V_{\rm atom}}}\frac{\sqrt{\chi_B}}{\chi_A}$, $d_2:=\frac{\IE_A\IE_B}{\exp{(V_{\rm cell}/\sum{V_{\rm atom}})}}$ & 99.6\%\textsuperscript{\ref{fn:YP}} \\
\hline
NaCl, CsCl, ZnS, CaF$_2$, Cr$_3$Si, SiC, TiO$_2$, ZnO, FeAs, NiAs, Al$_2$O$_3$, La$_2$O$_3$, Th$_3$P$_4$, ReO$_3$, ThH$_2$ & 299 & $\IE_A$, $\IE_B$, $\chi_A$, $\chi_B$, $x_A$, $x_B$, $V_{\rm cell}/\sum{V_{\rm atom}}$ & $d_1:=\frac{x_B}{\sum{V_{\rm atom}}/V_{\rm cell}}\frac{\IE_B\sqrt{\chi_B}}{\chi_A}$, $d_2:=\chi_A^2\left||1-2x_A|-x_A^2\frac{\chi_B}{\chi_A}\right|$ & 99.0\%\footnote{Three entry misclassified: YP-compound in NaCl-prototype; Th$_3$As$_4$- and La$_3$Te$_4$-compounds in Th$_3$P$_4$-prototype.}\\ \hline
\end{tabularx}
\end{table*}

It is critical to limit the redundant and unnecessary primary features in $\Phi_0$ to enhance computational performance (the size of features space $\Phi_n$ increases very fast with $\#\Phi_0$) and to
increase \SIS\ success rate: the higher \#subspace/$\#\Phi$ the higher the probability that \SIS\ subspaces contain the best models. 
Starting from an empty $\Phi_0$, few primary features are added. 
\SISSO\ is then applied  to identify the best model, with $\bm{\hat{H}}$ as operators space. 
If an appropriate quality of the model is not achieved (e.g., the number of correctly classified materials is lower than a desired threshold)), other primary features are added in $\Phi_0$ to check for improvements. 
Primary features preserved in $\Phi_0$ may become redundant or unnecessary on a later stage, e.g. when new ones are added. 
To retain computationally manageable sizes of the features space, tests are performed to remove those primary features that either are never appearing in the identified descriptor or that do not improve the performance of the model (in this specific case, when the number of correctly classified materials does not increase). 
Eventually, $\Phi_0$ will converge to the best possible small set of primary features, along with the best models that can be generated from it.

\noindent {\bf Data variety.}
The influence of data variety on the descriptors is investigated and Table~\ref{tb:TABLE1} shows how the metal-insulator classification descriptors depend on the prototypes of training materials.

The first calculation starts with a data set of all the available materials (132) in NaCl-prototype. 
The initial features space, $\Phi_0$, contains the primary features of all the 10 atomic parameters (Table~\ref{tb:TABLE1}),
and one structural parameter of interatomic distance $d_{AB}$ to capture the geometrical differences between the training rock-salt materials. 
\SISSO\ is then applied:
(1) $\Phi_3$  is constructed;
(2) the best descriptor is identified from $\Phi_3$ for classifying the metals and insulators with 100\% accuracy. 
The simple descriptor is shown in Table~\ref{tb:TABLE1}. 
It indicates that a rock-salt compound tends to become non metal when the large interatomic distance is decreased with the radius of species $A$. 

Next, the number of prototypes is increased to 5, for a total of 217 materials. 
However, with the previous $\Phi_0$ and calculation-settings, \SISSO\ fails to identify a descriptor having perfect classification (there are 7 points in the overlap-region between the metal and non metal domains). 
The non-optimal outcome indicates that the classification depends on primary features not yet considered.
First, $\Phi_0$ is slimmed by reducing its size to 7  --- $\EA_A$, $\EA_B$, $v_A$, and $v_B$ are removed --- without affecting the quality of the predictions (8 points in the overlap-region).
Second,  two new features $\CN_A$ and $\CN_B$ are added ($\#\Phi_0\rightarrow9$) to describe the different coordination environments of the prototypes.
\SISSO\ finds a 2D descriptor from the constructed $\Phi_3$ with 100\% classification, shown in Table~\ref{tb:TABLE1}. 
From the descriptor, the geometrical differences between training materials are captured by the two features of $d_{AB}$ and $\CN_B$: systems belonging to such 5 prototypes with large $d_{AB}$ and small $\CN_B$ tend to be non metals.

The number of prototypes is increased to 10, for a total of 260 materials. 
As shown in Table~\ref{tb:TABLE1}, with the previous $\#\Phi_0=9$, the identified best descriptors is 2D have 99.6\% classification (only one point, YP-compound in NaCl-prototype, is misclassified). 
Although the classification is excellent, the descriptor is complicate.
Searching for a simplification, new primary features of atomic composition $x_A$, $x_B$, and $V_{\rm cell}/\sum{V_{\rm atom}}$ are introduced to replace $r_{\rm covA}$, $r_{\rm covB}$, $d_{AB}$, $\CN_A$, and $\CN_B$, leading to $\#\Phi_0\rightarrow7$.
With the same training materials, \SISSO\ finds a much simple descriptor having the same accuracy of 99.6\% (YP-compound remains misclassified). 
This result shows that the choice of proper primary features leads to descriptors' simplification.

Finally, all the available 15 prototypes of binary materials (299) are considered and used with the 7 primary features in $\Phi_0$.
With a constructed $\Phi_3$ of size $10^8$, \SISSO\ identifies the best 2D descriptor with a classification accuracy of 99.0\% (three misclassified compounds: YP-compound in NaCl-prototype, Th$_3$As$_4$ and La$_3$Te$_4$ in Th$_3$P$_4$-prototype). 
When new information --- compounds and/or prototypes --- is added, the functional form of the descriptors adapts. 
For predictive models, the data set requires all necessary information, e.g., by uniform sampling of the whole chemical and configurational space of the property of interest. 
The above 15 prototypes are not all the available prototypes for binary materials, and the layered materials
(e.g., MoS$_2$, and those materials having $A-A$ or $B-B$ dimers, e.g., FeS$_2$, are not included) as the presented model is strictly illustrative of the method. \\

\noindent {\bf Reproducibility.}
To enable reproducibility, online tutorials where results can be interactively reproduced (and extended) are presented within the framework of the {\small NOMAD} Analytics-Toolkit ({\sf analytics-toolkit.nomad-coe.eu}). \\
For the RS/ZB benchmark application: \\ 
{\sf analytics-toolkit.nomad-coe.eu/tutorial-SIS.} \\
For the metal-nonmetal classification:\\
{\sf analytics-toolkit.nomad-coe.eu/tutorial-metal-nonmetal.}
The \SISSO\ code, as used for the work presented here, but ready for broader applications is open source and can be found at 
{\sf github.com/rouyang2017/SISSO.}





\newcommand{\Ozolins}{Ozoli\c{n}\v{s}}


\begin{thebibliography}{10}
\bibitem{SISSOnoteELLE0}
{T}he $\ell_0$-norm of a vector is the number of its non-zero components.
\input{SISSOnoteELLE0}
\bibitem{SISSOnoteELLE1}
{T}he $\ell_1$-norm is the sum of the absolute values of the components of a
  vector.
\input{SISSOnoteELLE1}
\end{thebibliography}
\end{document}